# GAMMA-400 gamma-ray observatory

**N.P. Topchiev[a1], A.M. Galper[a,b], V. Bonvicini[c], O. Adriani[d], R.L. Aptekar[e], I.V. Arkhangelskaja[b], A.I. Arkhangelskiy[b], A.V. Bakaldin[b], L. Bergstrom[f], E. Berti[d], G. Bigongiari[g], S.G. Bobkov[h], M. Boezio[c], E.A. Bogomolov[e], L. Bonechi[g], M. Bongi[d], S. Bottai[d], G. Castellini[i], P.W. Cattaneo[j], P. Cumani[c,p], O.D. Dalkarov[a], G.L. Dedenko[b], C. De Donato[k], V.A. Dogiel[a], N. Finetti[d], D. Gascon[l], M.S. Gorbunov[h], Yu.V. Gusakov[a], B.I. Hnatyk[m], V.V. Kadilin[b], V.A. Kaplin[b], A.A. Kaplun[b], M.D. Kheymits[b], V.E. Korepanov[n], J. Larsson[o], A.A. Leonov[a,b], V.A. Loginov[b], F. Longo[c], P. Maestro[g], P.S. Marrocchesi[g], M. Martinez[p], A.L. Men'shenin[q], V.V. Mikhailov[b], E. Mocchiutti[c], A.A. Moiseev[r], N. Mori[d], I.V. Moskalenko[s], P.Yu. Naumov[b], P. Papini[d], J.M. Paredes[l], M. Pearce[o], P. Picozza[k], A. Rappoldi[j], S. Ricciarini[i], M.F. Runtso[b], F. Ryde[o], O.V. Serdin[h], R. Sparvoli[k], P. Spillantini[d], Yu.I. Stozhkov[a], S.I. Suchkov[a], A.A. Taraskin[b], M. Tavani[t], A. Tiberio[d], E.M. Tyurin[b], M.V. Ulanov[e], A. Vacchi[c], E. Vannuccini[d], G.I. Vasilyev[e], J.E. Ward[p], Yu.T. Yurkin[b], N. Zampa[c], V.N. Zirakashvili[u], and V.G. Zverev[b]**

[1]Speaker, e-mail: *tnp51@yandex.ru*

[a] *Lebedev Physical Institute, RU-119991 Moscow, Russia*
[b] *National Research Nuclear University MEPhI, RU-115409 Moscow, Russia*
[c] *Istituto Nazionale di Fisica Nucleare, Sezione di Trieste, I-34149 Trieste, Italy*
[d] *Istituto Nazionale di Fisica Nucleare, Sezione di Florence, I-50019 Sesto Fiorentino, Florence, Italy*
[e] *Ioffe Physical Technical Institute, RU-194021 St. Petersburg, Russia*
[f] *The Oskar Klein Centre, Department of Physics, Stockholm University, AlbaNova University Center, SE-106 91 Stockholm, Sweden*
[g] *Department of Physical Sciences, Earth and Environment, University of Siena and Istituto Nazionale di Fisica Nucleare, Sezione di Pisa, Italy*
[h] *Scientific Research Institute for System Analysis, RU-117218 Moscow, Russia*
[i] *Istituto di Fisica Applicata Nello Carrara, I-50019 Sesto Fiorentino, Florence, Italy*
[j] *Istituto Nazionale di Fisica Nucleare, Sezione di Pavia, I-27100 Pavia, Italy*
[k] *Istituto Nazionale di Fisica Nucleare, Sezione di Rome "Tor Vergata", I-00133 Rome, Italy*
[l] *Departament d'Astronomia i Meteorologia, Institut de Ciències del Cosmos, Universitat de Barcelona, Spain*
[m] *Taras Shevchenko National University, Kyiv, 01601 Ukraine*
[n] *Lviv Center of Institute of Space Research, Lviv, 79060 Ukraine*
[o] *KTH Royal Institute of Technology, Department of Physics and the Oskar Klein Centre, AlbaNova University Center, SE-10691 Stockholm, Sweden*
[p] *Institut de Física d'Altes Energies, Bellaterra, Spain*
[q] *Research Institute for Electromechanics, RU-143502 Istra, Moscow region, Russia*
[r] *NASA Goddard Space Flight Center and CRESST/University of Maryland, Greenbelt, Maryland 20771, USA*
[s] *Hansen Experimental Physics Laboratory and Kavli Institute for Particle Astrophysics and Cosmology, Stanford University, Stanford, CA 94305, USA*
[t] *Istituto Nazionale di Astrofisica IASF and Physics Department of University of Rome Tor Vergata, I-00133 Roma, Italy*
[u] *Pushkov Institute of Terrestrial Magnetism, Ionosphere, and Radiowave Propagation, Troitsk, Moscow region, Russia*

The GAMMA-400 gamma-ray telescope with excellent angular and energy resolutions is designed to search for signatures of dark matter in the fluxes of gamma-ray emission and electrons + positrons. Precision investigations of gamma-ray emission from Galactic Center, Crab, Vela, Cygnus, Geminga, and other regions will be performed, as well as diffuse gamma-ray emission, along with measurements of high-energy electron + positron and nuclei fluxes. Furthermore, it will study gamma-ray bursts and gamma-ray emission from the Sun during periods of solar activity. The energy range of GAMMA-400 is expected to be from ~20 MeV up to TeV energies for gamma rays, up to 10 TeV for electrons + positrons, and up to $10^{15}$ eV for cosmic-ray nuclei. For high-energy gamma rays with energy from 10 to 100 GeV, the GAMMA-400 angular resolution improves from 0.1° to ~0.01° and energy resolution from 3% to ~1%; the proton rejection factor is ~$5 \times 10^5$. GAMMA-400 will be installed onboard the Russian space observatory.

## 1. Introduction

The main goal for the GAMMA-400 mission is to perform a sensitive search for signatures of dark matter particles in high-energy gamma-ray emission. This task was set for the GAMMA-400 project by Nobel Laureate Academician V.L. Ginzburg in the end of 1980’s [1, 2] and his list of very important issues in modern cosmology at the beginning of XXI Century noted the issue of dark matter and its detection [3]. Within the framework of this project, which has now since expanded internationally, the design and construction of a future, complex, precision gamma-ray telescope is being carried out [4-7].

It should be noted, up to now, neither in indirect measurements of EGRET [8], currently operating Fermi-LAT [9], ground-based gamma-ray facilities [10] nor in direct measurements of various ground-based experiments including experiments at LHC, any direct evidence of hypothetical dark matter particles were not found.

GAMMA-400 is designed to search for signatures of dark matter in the fluxes of gamma-ray emission and electrons + positrons. Precision investigations of gamma-ray emission from Galactic Center, Crab, Vela, Cygnus, Geminga, and other regions will be performed, as well as diffuse gamma-ray emission, along with measurements of high-energy electron + positron and nuclei fluxes. Furthermore, it will study gamma-ray bursts and gamma-ray emission from the Sun during periods of solar activity.

This paper presents the basic physical and technical characteristics of GAMMA-400 and the measurement conditions.

## 2. The GAMMA-400 gamma-ray telescope

The GAMMA-400 physical scheme is shown in Fig. 1. GAMMA-400 consists of plastic scintillation anticoincidence top and lateral detectors (ACtop and AClat), a converter-tracker (C), plastic scintillation detectors (S1 and S2) for a time-of-flight system (ToF), a two-part calorimeter (CC1 and CC2), lateral detectors (LD), plastic scintillation detectors (S3 and S4), and a neutron detector (ND).

The anticoincidence detectors surrounding the converter-tracker are used to distinguish gamma rays from significantly larger number of charged particles (e.g., in the region of 10-100 GeV, the flux ratios for gamma rays to electrons and protons are ~ $1:10^2:10^5$ [11]).

All scintillation detectors consist from two independent layers, with each 1-cm layer. The time-of-flight system, where detectors S1 and S2 are separated by approximately 500 mm, determines the top-down direction of arriving particles. The additional scintillation detectors S3 and S4 improve hadron and electromagnetic showers separation.

The converter-tracker consists of 13 layers of double (x, y) silicon strip coordinate detectors (pitch of 0.08 mm). The first three and final two layers have no tungsten while the middle eight layers are interleaved with tungsten conversion foils. Using the first three layers without tungsten allows us to measure gamma rays down to approximately 20 MeV. The total converter-tracker thickness is about 1 $X_0$ (where $X_0$ is the radiation length). The converter-tracker information is utilized to precisely determine the conversion point and the direction of each incident particle.

The two-part calorimeter measures particle energy. The imaging calorimeter CC1

consists of 2 layers of double (x, y) silicon strip coordinate detectors (pitch of 0.08 mm) interleaved with planes from CsI(Tl) crystals, and the electromagnetic calorimeter CC2 consists of CsI(Tl) cubic crystals with dimensions of 36 mm × 36 mm × 36 mm. The thickness of CC1 and CC2 is 2 $X_0$ and 23 $X_0$, respectively. The total calorimeter thickness is 25 $X_0$ or 1.2 $\lambda_0$ (where $\lambda_0$ is nuclear interaction length) when detecting vertical incident particles and 54 $X_0$ or 2.5 $\lambda_0$ when detecting laterally incident particles. Using a deep calorimeter allows us to extend the energy range up to several TeV for gamma rays, 10 TeV for electrons, and to reach an energy resolution of approximately 1% above 100 GeV.

Neutron detector is used to separate hadron and electromagnetic showers.

The GAMMA-400 principal possibilities to detect particles from vertical and lateral directions are shown in Fig. 1 for several fields of view: (1) for gamma rays and electrons + positrons with the full aperture of ±60° and with the best angular resolution of ~0.01° within ±30° from vertical ($E_\gamma$ = 100 GeV), (2) for gamma-rays and electrons + positrons with the angular resolution of ~0.2° ($E_\gamma$ = 100 GeV), and (3) for gamma-rays, electrons + positrons, and nuclei with the angular resolution of ~3° ($E_\gamma$ = 100 GeV).

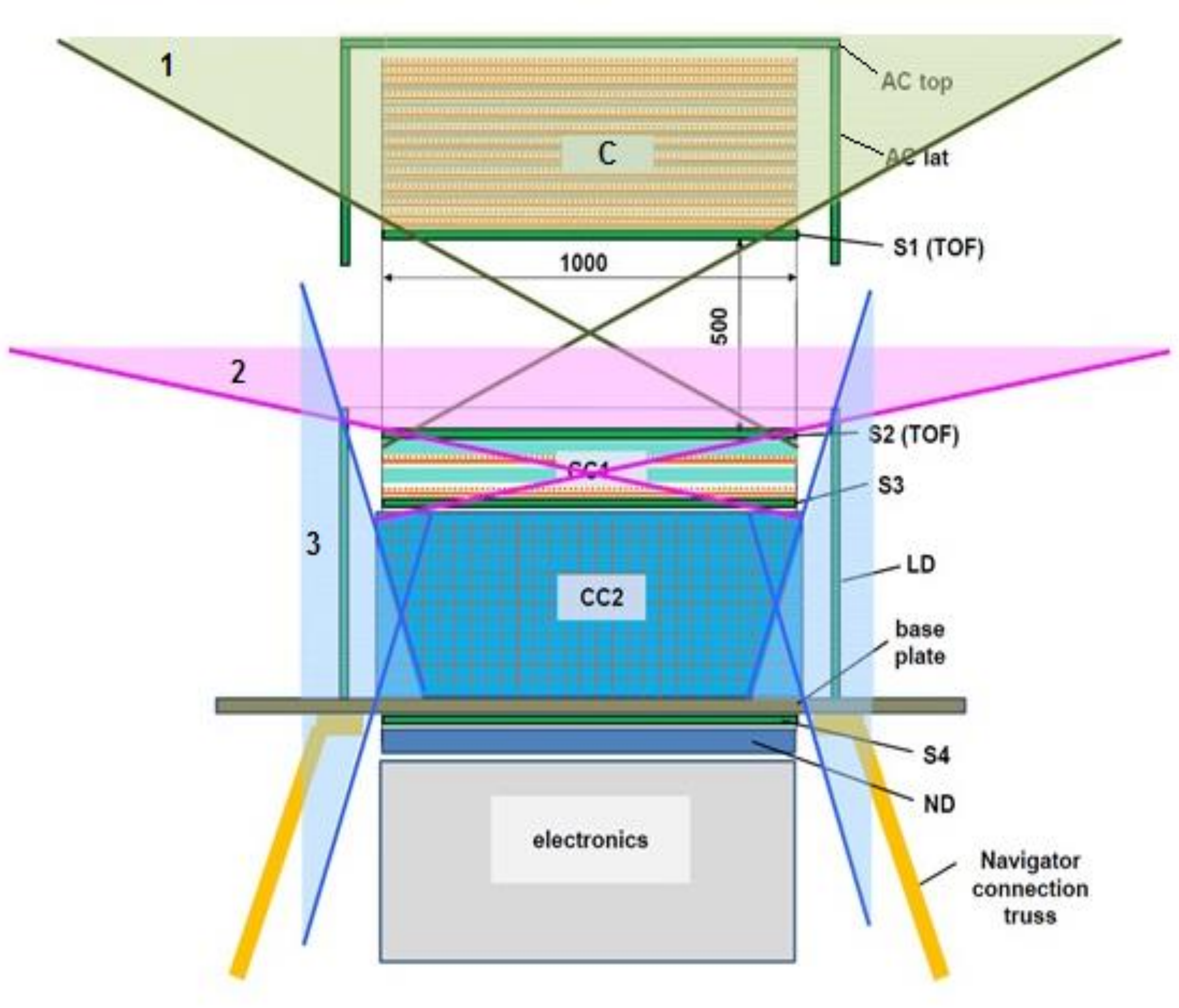

**Figure 1**: The GAMMA-400 physical scheme.

## 3. The GAMMA-400 performance and comparison with Fermi-LAT and ground-based facilities

The GAMMA-400 performance has been presented earlier in [4-7]. At present, using additional first three double layers (with no tungsten) allows us to achieve the best angular resolution in the energy range from ~20 MeV up to TeV in spite of some improved the Fermi-LAT performance with Pass 8 (http://www.slac.stanford.edu/exp/glast/groups/canda/lat_Performance.htm).

Table 1 presents a comparison of the Fermi-LAT and GAMMA-400 parameters. GAMMA-400 has numerous advantages:

- highly elliptical orbit (without the Earth's occultation and away from the radiation belts) allows us to observe with the full aperture of ±60° different gamma-ray sources continuously over a long period of time with the exposition greater by a factor of 7 than for Fermi-LAT operating in the sky-survey mode;

- thanks to a smaller pitch (by a factor of 3) and analog readout in the coordinate silicon strip detectors, GAMMA-400 has an excellent angular resolution above ~20 MeV;

- due to the deep (~25 $X_0$) calorimeter, GAMMA-400 has an excellent energy resolution and can more reliably to detect gamma rays up to several TeV and electrons + positrons up to 10 GeV for vertically incident events. When measuring lateral events (electrons + positrons and nuclei) it can detect diffuse gamma-ray emission and cosmic rays using only the CC2 calorimeter;

- owing to the better gamma-ray separation from cosmic rays (in contrast to Fermi-LAT, the presence of a special trigger with event timing, time-of-flight system, two-layer scintillation detectors), GAMMA-400 is significantly well equipped to separate gamma rays from the background of cosmic rays and backscattering events.

GAMMA-400 will have also the better angular and energy resolutions in the energy region 10-1000 GeV in comparison with current and future space- and ground-based instruments: H.E.S.S. [12], MAGIC [13], VERITAS [14], HAWC [15], CTA [16], DAMPE [17], and CALET [18] (Fig. 2, Table 2) and it allows us to fill the gap between the space- and ground-based instruments. It should be noted that there is a preliminary agreement between the GAMMA-400 and CTA leaders about future joint simultaneous observations.

GAMMA-400 will study continuously over a long period of time different regions of Galatic plane, for example, Galactic Center, Crab, Vela, Cygnus, Geminga with the full aperture of ±60°. In particular, using the gamma-ray fluxes obtained by Fermi-LAT [19] and http://fermi.gsfc.nasa.gov/ssc/data/access/, we can expect that GAMMA-400 when observing the Galactic Center with aperture of ±45°during 1 year will detect:

57400 photons for $E_\gamma > 10$ GeV;

5240 photons for $E_\gamma > 50$ GeV;

1280 photons for $E_\gamma > 100$ GeV;

535 photons for $E_\gamma > 200$ GeV.

Table 1

| | **Fermi-LAT** | **GAMMA-400** |
|---|---|---|
| Orbit | Circular, 565 km | Highly elliptical, 500-300000 km (without the Earth's occultation) |
| Operation mode | Sky-survey (3 hours) | Point observation (up to 100 days) |
| Source exposition | 1/7 | 1 |
| Energy range | 20 MeV - 300 GeV (γ, e) | ~20 MeV - 1 TeV (γ) 1 GeV - 10 TeV (e) |
| Effective area ($E_\gamma$ > 1 GeV) | ~6500 $cm^2$ (total) ~4000 $cm^2$ (front) | ~4000 $cm^2$ |
| Coordinate detectors - readout | Si strips (pitch 0.23 mm) digital | Si strips (pitch 0.08 mm) analog |
| Angular resolution | ~4° ($E_\gamma$ = 100 MeV) ~0.2° ($E_\gamma$ = 10 GeV) ~0.1° ($E_\gamma$ > 100 GeV) | ~2° ($E_\gamma$ = 100 MeV) ~0.1° ($E_\gamma$ = 10 GeV) ~0.01° ($E_\gamma$ > 100 GeV) |
| Calorimeter - thickness | CsI(Tl) ~8.5 $X_0$ | CsI(Tl) + Si ~25 $X_0$ |
| Energy resolution | ~10% ($E_\gamma$ = 10 GeV) ~10% ($E_\gamma$ > 100 GeV) | ~3% ($E_\gamma$ = 10 GeV) ~1% ($E_\gamma$ > 100 GeV) |
| Proton rejection factor | ~$10^3$ | ~$5x10^5$ |
| Mass, kg | 2800 | 4100 |
| Telemetry downlink volume, Gbytes/day | 15 | 100 |

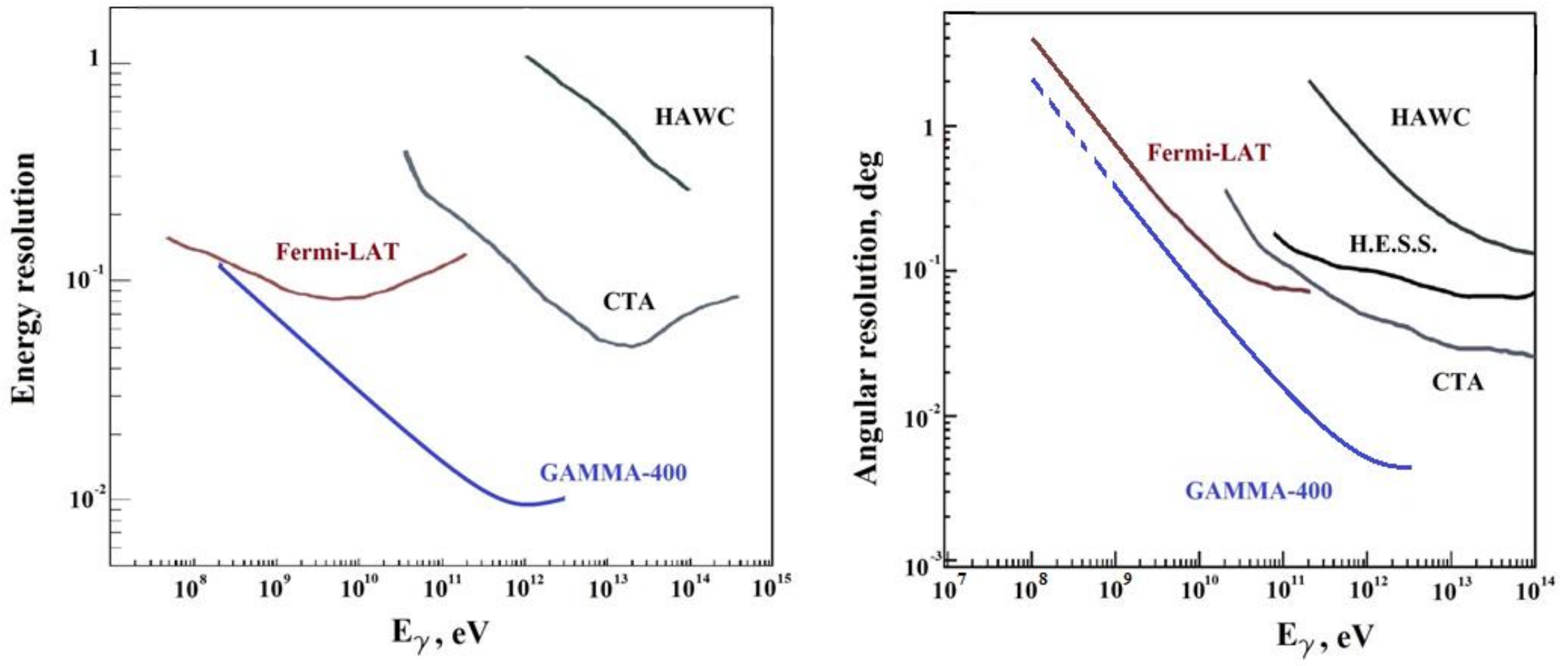

**Figure 2:** Comparison of energy and angular resolutions for GAMMA-400, Fermi-LAT, H.E.S.S., HAWC, and CTA (Fig. 2 from [20] was used).

Table 2

| | **SPACE-BASED GAMMA-RAY INSTRUMENTS** | | | | **GROUND-BASED GAMMA-RAY INSTRUMENTS** | | | |
|---|---|---|---|---|---|---|---|---|
| | **Fermi-LAT** | **DAMPE** | **CALET** | **GAMMA-400** | **H.E.S.S.** | **MAGIC** | **VERITAS** | **CTA** |
| Particles | γ, e | e, nuclei, γ | e, nuclei, γ | γ, e, nuclei | γ | γ | γ | γ |
| Operation period | 2008- | 2015 | 2015 | ~2023 | 2012- | 2009- | 2007- | ~2020 |
| Energy range, GeV | 0.02-300 | 5-10000 | 10-10000 | **0.02-10000** | > 30 | > 50 | > 100 | > 20 |
| Angular resolution ($E_\gamma$ > 100 GeV) | 0.1° | 0.1° | 0.1° | ~**0.01**° | 0.07° | 0.07° ($E_\gamma$ = 300 GeV) | 0.1° | 0.1° ($E_\gamma$ = 100 GeV) 0.03° ($E_\gamma$ = 10 TeV) |
| Energy resolution ($E_\gamma$ > 100 GeV) | 10% | 1.5% | 2% | ~**1%** | 15% | 20% ($E_\gamma$ = 100 GeV) 15% ($E_\gamma$ = 1 TeV) | 15% | 20% ($E_\gamma$ = 100 GeV) 5% ($E_\gamma$ = 10 TeV) |

## 4. GAMMA-400 gamma-ray observatory

In addition to the gamma-ray telescope, the GAMMA-400 scientific complex includes the KONUS-FG gamma-ray burst monitor and two star sensors for determining the gamma-ray telescope axis with an accuracy of ~5″, along with two magnetometers for measuring the magnetic field. The GAMMA-400 gamma-ray observatory will be installed onboard of the Navigator space platform, which is designed and manufactured by the Lavochkin Association.

Using the Navigator space platform gives the GAMMA-400 experiment a highly unique opportunity for the near future gamma- and cosmic-ray science, since it allows us to install a very large scientific payload (a mass of 4100 kg, power consumption of 2000 W, and a telemetry downlink of 100 GB/day, with lifetime more than 7 years), which will provide GAMMA-400 with the means to significantly contribute as the next generation instrument for gamma-ray astronomy and cosmic-ray physics.

The GAMMA-400 experiment will be initially launched into a highly elliptical orbit (with an apogee of 300,000 km and a perigee of 500 km, with an inclination of 51.4°), with 7 days orbital period. Under the action of gravitational disturbances of the Sun, Moon, and the Earth after ~6 months the orbit will transform to about an approximately circular one with a radius of ~200,000 km and will not suffer from the Earth’s occultation and shielding by the radiation belts. A great advantage of such an orbit is the fact that the full sky coverage will always be available for gamma-ray astronomy, since the Earth will not cover a significant fraction of the sky, as is usually the case for low Earth orbit. Therefore, the GAMMA-400 source pointing strategy will hence be properly defined to maximize the physics outcome of the experiment. The launch of the GAMMA-400 space observatory is planned for the middle of the 2020s.